\documentclass[12pt]{article}

\newcommand{\sect}[1]{\setcounter{equation}{0}\section{#1}}

\newcommand{\eq}{\begin{equation}}
\newcommand{\eqa}{\begin{eqnarray}}
\newcommand{\en}{\end{equation}}
\newcommand{\ena}{\end{eqnarray}}
\newcommand{\enn}{\nonumber \end{equation}}

\def\sk{\vskip .4cm}
\def\noi{\noindent}
\def\om{\omega}
\def\al{\alpha}
\def\be{\beta}
\def\ga{\gamma}
\def\Ga{\Gamma}

\let \si\sigma
\let \part\partial

\def\unmezzo{{1 \over 2}}
\def\epsi{\varepsilon}
\def\we{\wedge}

\def\de{\delta}

\def\tv{{\bf t}}
\def\vbo{{\bf v}}
\def\Gt{{\tilde G}}

\def\part{\partial}

\def\sk{\vskip .4cm}

\def\noi{\noindent}

\def\X0{X^0}

\def\om{\omega}

\def\al{\alpha}
\def\ga{\gamma}

\def\unmezzo{{1 \over 2}}
\def\epsi{\varepsilon}
\def\epsibo{{\bf  \epsilon}}
\def\epsib{{\bar \epsi}}
\def\psib{{\bar \psi}}

\def\rhob{\bar\rho}

\def\we{\wedge}

\def\de{\delta}
\def\CABC{{C^A}_{BC}}

\def\tbo{{\bf t}}
\def\Gt{{\tilde G}}

\def\c#1#2{ C_{~#1}^{#2} }
\def\cl#1#2{ C_{~#1}^{#2} }

\def\Dcal{{\cal D}}

\def\square{{\,\lower0.9pt\vbox{\hrule \hbox{\vrule height 0.2 cm
\hskip 0.2 cm \vrule height 0.2 cm}\hrule}\,}}

\begin{document}

\begin{titlepage}
\rightline{DISTA-UPO/05}
\rightline{August 2005}
 \vskip 2em
\begin{center}{\bf   LIE DERIVATIVES ALONG ANTISYMMETRIC TENSORS, AND THE M-THEORY
SUPERALGEBRA}
\\[3em]
Leonardo Castellani\\[1em] {\sl Dipartimento di Scienze e
Tecnologie Avanzate\\ Universit\`a del Piemonte Orientale\\ Via
Bellini 25/G, 15100 Alessandria, Italy}\\[.5 em] {\sl and\\[.5em]
Istituto Nazionale di Fisica Nucleare\\ Via Giuria 1, 10125
Torino, Italy.}
  \\[3em]

\end{center}

\begin{abstract}
Free differential algebras (FDA's) provide an algebraic setting
for field theories with antisymmetric tensors. The ``presentation"
of FDA's generalizes the Cartan-Maurer equations of ordinary Lie
algebras, by incorporating $p$-form potentials. An {\sl extended
Lie derivative} along antisymmetric tensor fields can be defined,
and used to recover a Lie algebra dual to the FDA, that encodes
all the symmetries of the theory {\sl including those
 gauged by the $p$-forms}.

 The general method is applied to the FDA of $D=11$ supergravity: the
 resulting dual Lie superalgebra contains the M-theory supersymmetry anticommutators
 in presence of 2-branes.

\end{abstract}

\vskip 6cm \noi \hrule \vskip.2cm \noi {\small
leonardo.castellani@mfn.unipmn.it}

\end{titlepage}

\newpage
\setcounter{page}{1}

\sect{Introduction}

Supergravity in eleven dimensions \cite{d11} is today considered
an effective theory (a particular limit of M-theory, for a review
see for ex. \cite{mtheory}). More than two decades ago, it was
formulated \cite{fda1} as the gauging of a free differential
algebra (FDA) \cite{fda0,fda1,fda2,gm2}, an algebraic structure
that extends the Cartan-Maurer equations of an ordinary Lie
algebra $G$ by including $p$-form potentials, besides the usual
left-invariant one-forms corresponding to the Lie group generators
of $G$. Thus the 3-form of D=11 supergravity acquires an algebraic
interpretation, as well as the $p$-forms present in supergravity
theories in various dimensions.

The group-geometric method of \cite{gm1,gm2,gm3} yields
lagrangians based on given FDA's. These FDA's encode the
symmetries of the resulting field theories.

Only some time later it was realized how to extract from the FDA
also the symmetries gauged by the $p$-forms, via a new
(``extended") Lie derivative defined along antisymmetric tensors
\cite{fda3}. The extended Lie derivatives, together with the
ordinary Lie derivatives of the $G$ Lie algebra contained in the
FDA,  close on an algebra that can be considered dual to the FDA.

The transformations on the fields generated by the extended Lie
derivatives are the symmetries gauged by the antisymmetric
tensors, and can be explicitly computed.

In this paper we generalize the treatment of \cite{fda3} (limited
to 2-forms) to include arbitrary $p$-forms, and apply it to the
FDA of D=11 supergravity. The resulting dual Lie superalgebra
contains the supersymmetry anticommutators of M-theory coupled to
a 2-brane discussed in \cite{2brane} , one of the extended Lie
derivatives corresponding to the pseudo-central charge
$Z^{m_1m_2}$.

In fact a supertranslation algebra containing pseudo-central
charges $Z^{m_1m_2}$ and $Z^{m_1-m_5}$ had already been found by
D'Auria and Fr\'e, who proposed in \cite{fda1} a method to
``resolve" FDA's into ordinary Lie algebras by considering the
$p$-forms as composites of 1-form potentials of a larger group,
containing the generators of $G$ plus some extra generators. For
the FDA of $D=11$ supergravity, the extra generators were found to
be the two pseudo-central charges $Z^{m_1m_2}$ and $Z^{m_1-m_5}$,
and an additional spinorial charge $Q'$.

Here we obtain a similar (but not identical) algebra: besides
$Z^{m_1m_2}$ we find a vector-spinor charge $Q^m$.

Closer contact with the D'Auria-Fr\'e algebra can be achieved by
further extending our treatment to FDA's containing more than one
$p$-form. Then we can apply it to a FDA containing a 3-form and a
6-form, so that both the charges $Z^{M_1M_2}$ and $Z^{M_1-M_5}$
enter the stage in the dual Lie algebra. This leads to the same
supertranslation algebra of \cite{fda1}, that later was derived
\cite{2and5brane} in the context of D=11 supergravity coupled to a
2- and a 5-brane.

A resum\'e on FDA's and their gauging is given in Section 2. By
use of the extended Lie derivatives we obtain the dual formulation
of FDA's containing a $p$-form. Both the soft and rigid FDA
diffeomorphism algebras are given (the latter being a Lie algebra
for constant parameters). This is applied in Section 3 to the FDA
of D=11 supergravity. In Section 4 we discuss the possibility of
gauging the superalgebra dual of this FDA, thus obtaining a new
formulation of $D=11$ supergravity.

\sect{Free differential algebras and their Lie algebra duals}

Rather than the general theory of FDA's (for a detailed review see
\cite{gm2}, and \cite{gm3} for a shorter account), we'll treat
here the case involving only one $p$-form. It already contains
most of the essential features of FDA's. Its ``presentation" is
given by the generalized Cartan-Maurer equations: \eqa & & d
\sigma^A + \unmezzo \CABC ~\sigma^B \sigma^C = 0 \label{fdaLie}\\
& & dB^i+C^i_{~Aj} \sigma^A B^j + {1 \over (p+1)!}
C^i_{~A_1...A_{p+1}} \sigma^{A_1}... \sigma^{A_{p+1}}\nonumber \\
 & & ~~~~~~~~~~~~~\equiv \nabla B^i +  {1 \over
(p+1)!} C^i_{~A_1...A_{p+1}} \sigma^{A_1}... \sigma^{A_{p+1}} =0
\label{fdaB} \ena

\noi where $\sigma^A$ are the usual left-invariant one-forms
associated to a Lie algebra $G$, $B^i$ is a $p$-form in a
representation $D^i_{~j}$ of $G$, and products between forms are
understood to be exterior products.

 The Jacobi identities for the
generalized structure constants, ensuring the integrability of
(\ref{fdaLie}),(\ref{fdaB}), i.e. the nilpotency of the external
derivative $d^2=0$, are:
\eqa & & C^A_{~~B[C} C^B_{~~DE]} =0
\label{jacobi1} \\
   & &
C^i_{~Aj} C^j_{~Bk} - C^i_{~Bj} C^j_{~Ak} = C^C_{~AB} C^i_{~Ck}
\label{jacobi2} \\ & & 2 ~C^i_{~[A_1j} C^j_{~A_2...A_{p+2}]}
-(p+1) C^i_{~B[A_1...A_p} C^B_{~A_{p+1}A_{p+2}]}=0
 \label{jacobi3} \ena

\noi Eq. (\ref{jacobi1}) are the usual Jacobi identities for the
Lie algebra $G$. Eq. (\ref{jacobi2}) implies that $(C_A)^i_{~j}
\equiv C^i_{~Aj}$ is a matrix representation of $G$, while eq.
(\ref{jacobi3}) states that $C^i \equiv C^i_{~A_1...A_{p+1}}
\sigma^{A_1}... \sigma^{A_{p+1}}$ is a $(p+1)$-cocycle, i.e.
$\nabla C^i = 0$.

\sk

\subsection{Dynamical fields, curvatures and Bianchi identities}
\sk

The main idea of the group-geometric method  \cite{gm1,gm2}
extended to FDA's is to consider the one-forms $\si^A$ and the
$p$-form $B^i$ as the fundamental fields of the geometric theory
to be constructed. In the case of ordinary Lie algebras the
dynamical fields are the vielbeins $\mu^A$ of $\Gt$, a smooth
deformation of the group manifold $G$ referred to as ``soft group
manifold". For FDA's the dynamical fields are both the  vielbeins
$\mu^A$ and the $p$-form field $B^i$: taken together they can be
considered the vielbeins of the ``soft FDA manifold".

In general $\mu^A$ and $B^i$ do not satisfy any more the
Cartan-Maurer equations (\ref{fdaLie}),(\ref{fdaB}), so that
\eqa
& & R^A \equiv d\mu^A+{1 \over 2} \CABC \mu^B \we \mu^C \not= 0
\label{RLie}\\
 & & R^i = dB^i+C^i_{~Aj} \mu^A B^j + {1 \over
(p+1)!} C^i_{~A_1...A_{p+1}} \mu^{A_1}... \mu^{A_{p+1}}  \not= 0
\label{RFDA}
 \ena

 \noi  The extent of the deformation of the
FDA is measured by the curvatures: the two-form $R^A$ and the
$(p+1)$-form $R^i$. (Note: we use the same symbol $B^i$ for the
``flat" and the ``soft" $p$-form). The deformation of the FDA is
necessary in order to allow field configurations with nonvanishing
curvatures.
 \sk
 Applying the external derivative $d$ to the definition of $R^A$
and $R^i$ (\ref{RLie}),(\ref{RFDA}), using $d^2 =0$ and the Jacobi
identities (\ref{jacobi1})-(\ref{jacobi3}), yields the Bianchi
identities :
\eqa
       & & dR^A-\CABC~ R^B \mu^C=0 \label{BianchiRA}\\
       & & dR^i-C^i_{~Aj} R^A B^j+C^i_{~Aj} \mu^A R^j - {1 \over p!}
        C^i_{~A_1...A_{p+1}} R^{A_1} \mu^{A_2}... \mu^{A_{p+1}}=0 \label{BianchiRi}
\ena

\noi The curvatures can be expanded on the $\mu^A, B^i$ basis of
the ``soft FDA manifold" as
 \eqa
  & & R^A={R^A}_{BC} \mu^B \mu^C+R^A_{~~i } B^i \label{RAexpansion}
  \\
   & & R^i=R^i_{~A_1...A_{p+1}} \mu^{A_1}... \mu^{A_{p+1}}  + R^i_{~Aj} \mu^A B^j
\label{Riexpansion}
  \ena

  \noi (Note: the $R^A_{~~i } B^i$ term in (\ref{RAexpansion}) can be there only for
  $p=2$).
  The FDA vielbeins $\mu^A$ and $B^i$ are a basis for the FDA ``manifold".
  Coordinates $y$ for this ``manifold" run on the corresponding ``directions", i.e. Lie algebra directions
  and ``$p$-form directions". The coordinates running on the $p$-form directions
  are $p-1$ forms (generalizing the coordinates running on the Lie algebra
  directions, which are 0-forms).

 Eventually we want {\sl space-time} fields :  the only
 coordinates  the fields must depend on are spacetime coordinates, associated
 with the (bosonic) translation part of the algebra.
 This is achieved when the curvatures are {\sl
horizontal} in the other directions (see later).

How do we find the dynamics of $\mu^A (y)$ and $B^i(y)$ ? We wish
to obtain a geometric theory, i.e. invariant under
diffeomorphisms. We need therefore to construct an action
invariant under diffeomorphisms, and this is simply achieved by
using only diffeomorphic invariant operations as the exterior
derivative and the exterior product. The building blocks are the
one-form $\mu^A$ and the $p$-form $B^i$, their curvatures $R^A$
and $R^i$: exterior products of them can make up a lagrangian
D-form, where D is the dimension of space-time.

A detailed account of the procedure, together with various
examples of supergravity theories based on FDA's, can be found in
\cite{gm2,gm3}.

\subsection{ Diffeomorphisms and Lie derivative}

The variation under diffeomorphisms $y + \epsi$ of an arbitrary
form $\om(y)$ on a manifold is given by the Lie derivative of the
form along the infinitesimal tangent vector $\epsilon=\epsi^M
\part_M$ :
\eq \de \om = \om (y+\epsi)-\om(y)=d(i_{\epsilon}\om)
       +i_{\epsilon} d \om \equiv \ell_{\epsilon} \om \label{diff}
\en

\noi On $p$-forms $\om_{(p)}=\om_{M_1...M_p}dy^{M_1} \we ... \we
dy^{M_p} $, the {\sl contraction} $i_{\bf v}$ along an arbitrary
tangent vector ${\bf v}= v^M  \part_M$ is defined as \eq i_{\bf
v}~ \om_{(p)}=p~ v^{M_1} \om_{M_1M_2...M_p}~dy^{M_2} \we ... \we
dy^{M_p}
 \label{contraction}
\en
and maps $p$-forms into ($p-1$)-forms. On the vielbein basis eq.
(\ref{contraction}) becomes \eq i_{\bf v}~ \om_{(p)}=p~ v^A
\om_{AB_2...B_p}~\mu^{B_2}
 \we ... \we \mu^{B_p}
    \label{contraction2}
\en

\noi where as usual curved indices {\small {\sl (M,N,...)}} are
related to tangent indices {\small {\sl (A,B,...)}} via the
vielbein (or inverse vielbein) components $\mu^A_M$ ($\mu^M_A$),
i.e. ${\bf v}= v^M
\part_M =
  v^A \tbo_A$ where $\tbo_A \equiv \mu^M_A \part_M$ etc.
  Thus the tangent vectors $\tbo_A$ are dual to the vielbeins:
  $\mu^B(\tbo_A)=\de^B_A$.

The operator \eq \ell_{\vbo} \equiv d~ i_{\vbo} + i_{\vbo} ~d
\label{Liederivative}
\en

\noi is the {\sl Lie derivative} along the tangent vector $\vbo$
and maps $p$-forms into $p$-forms.

\sk In the case of a group manifold $G$, we can rewrite the
vielbein variation under diffeomorphisms  in a suggestive way:
\eqa & & \de \mu^A = d(i_{\epsilon}\mu^A)
       +i_{\epsilon} d \mu^A = d \epsi^A  + 2  (d\mu^A)_{BC}~ \epsi^B
       \mu^C
  \nonumber \\ & &= (\nabla \epsi)^A + i_{\epsibo} R^A
      \label{diffgroup}
\ena

\noi where we have used the definition (\ref{RLie}) for the
curvature, and the $G$-covariant derivative $\nabla$ acts on
$\epsi^A$ as
 \eq (\nabla \epsi)^A \equiv d \epsi^A + \CABC \mu^B
\epsi^A \label{covdev}
\en
\sk

When dealing with FDA's,  what is the action of diffeomorphisms on
the $p$-form $B^i$ ? First, we consider diffeomorphisms in the Lie
algebra directions. For these, the Lie derivative formula
(\ref{diff}) holds. We have therefore, with tangent indices:
\eq
\de B^i = \ell_{\epsi^A \tbo_A} B^i = d(i_{\epsi^A \tbo_A}B^i)
       +i_{\epsi^A \tbo_A} d B^i \label{diffB1}
\en

\noi Since $\mu^A$ and $B^i$ are a basis for the  FDA ``manifold",
the contraction of $B^i$ along a Lie algebra tangent vector
$\tbo_A$ vanishes:
\eq i_{\tbo_A} \mu^B=\de^B_A,~~~i_{\tbo_A}B^i=0
\en

\noi and using the definition of $R^i$ (\ref{RFDA}) the variation
(\ref{diffB1}) takes the form:

\eqa
 & &\de B^i = i_{\epsi^A \tbo_A} d B^i =
 \nonumber \\
 & &=\left(R^i_{Aj} - C^i_{Aj}\right)\epsi^AB^j +
\left((p+1)R^i_{AA_1...A_p}
-\frac{1}{p!}C^i_{AA_1...A_p}\right)\epsi^A\mu^{A_1}...\mu^{A_p}\\
 & & \equiv (\nabla \epsi)^i + i_{\epsi^A \tbo_A} R^i
 \label{diffB2}
 \ena

 \subsection{Extended Lie derivatives }
 \sk

 Before
computing the algebra of Lie derivatives on the FDA fields, we
introduce
 \sk
  i) a new contraction operator $i_{\epsi^j \tv_j}$,
defined by its action on a generic form $\om = \om_{i_1...i_n
A_1...A_m} B^{i_1} \we ... B^{i_n} \we \mu^{A_1} \we ...
\mu^{A_m}$ as
\eq i_{\epsi^j \tv_j} \om = n~ \epsi^j \om_{j
i_2...i_n A_1...A_m} B^{i_2} \we ... B^{i_n} \we \mu^{A_1} \we ...
\mu^{A_m} \label{newcontraction}
\en
\noi {\sl where} $\epsi^j$ {\sl is a (p-1)-form}. This operator
still maps $p$-forms into $(p-1)$-forms. We can also define the
contraction $i_{\tv_j}$, mapping $n$-forms into $(n-p)$-forms, by
setting \eq i_{\epsi^j \tv_j}=\epsi^j i_{\tv_j}
\en
\noi In particular
 \eq
 i_{\tv_j} B^i=\de^i_j,~~~i_{\tbo_j} \mu^A =0
\en

\noi so that $\tv_j$ can be seen as the ``tangent vector" dual to
$B^j$. Note that $i_{\epsi^j \tv_j}$ vanishes on forms that do not
contain at least one factor $B^i$. \sk ii) a new Lie derivative
(``extended Lie derivative") given by:
 \eq
 \ell_{\epsi^i \tv_i} \equiv i_{\epsi^i \tv_i}d + d~
i_{\epsi^i \tv_i} \label{newLie}
 \en
The extended Lie derivative commutes with $d$, satisfies the
Leibnitz rule, and can be verified to act on the fundamental
fields as
 \eqa &
&\ell_{\epsi^j \tv_j}\mu^A = \epsi^j R^A_{~j} \label{newLieonmu}\\
& &\ell_{\epsi^j \tv_j}B^i = d\epsi^i + (C^i_{~Aj} - R^i_{~Aj})
\mu^A \we \epsi^j \label{newLieonB} \ena

\noi by applying the definitions of the curvatures  (\ref{RLie})
and (\ref{RFDA}).

\subsection{The algebra of diffeomorphisms}

Using the Bianchi identities (\ref{BianchiRA}), (\ref{BianchiRi}),
we find that the Lie derivatives {\sl and} the extended Lie
derivatives close on the algebra:
 \eqa & &
 \left[ \ell_{ \epsi^A_1 \tbo_A},\ell_{
\epsi^B_2 \tbo_B} \right] = \ell_{ \left[ \epsi^A_1 \partial_A
\epsi^C_2 - \epsi^A_2 \partial_A \epsi^C_1 +\epsi^A_1 \epsi^B_2
\left( C^C_{AB} -2R^C_{AB}\right) \right] \tbo_C} \nonumber \\
 & &~~~~~~~~~~~~~~~~~
 +\ell_{ 2 \epsi^A_1 \epsi^B_2~ ({1 \over p!}C^i_{ABA_1...A_{p-1}}
 - ~ R^i_{ABA_1...A_{p-1}})~\mu^{A_1}...\mu^{A_{p-1}}\tbo_i }
\label{FDAalgebra1}\\
  & &\left[ \ell_{ \epsi^A
\tbo_A} , \ell_{ \epsi^j \tbo_j} \right] = \ell_{[ \ell_{ \epsi^A
\tbo_A} \epsi^k + \left( C^k_{Bj} - R^k_{Bj} \right) \epsi^B
\epsi^j ] \tbo_k}\label{FDAalgebra2}\\ & &\left[ \ell_{ \epsi^i_1
\tbo_i} , \ell_{ \epsi^j_2 \tbo_j} \right] = \ell_{R^B_{~i}
(\epsi^i_1 (\epsi_2)^j_B - \epsi_2^i (\epsi_1)^j_B) \tbo_j}
\label{FDAalgebra3}\ena

\noi The last commutator between extended derivatives vanishes
except in the case $p=2$ (since only in this case $R^B_{~i}$ can
be different from 0: then $\epsi^i_A$ are the components of the
1-form $\epsi^i$, i.e. $\epsi^i \equiv \epsi^i_A \mu^A$).

 Notice that the commutator of two ordinary Lie derivatives {\sl contains
 an extra piece proportional to an extended Lie derivative}. This result has an important
consequence: if the field theory based on the FDA is geometric,
i.e. its action is invariant under diffeomorphisms generated by
usual Lie derivatives, then {\sl also the extended Lie derivative
must generate a symmetry of the action}, since it appears on the
right-hand side of (\ref{FDAalgebra1}). Thus, when we construct
geometric lagrangians gauging the FDA, we know {\sl a priori }
that the resulting theory will have symmetries generated by the
extended Lie derivative: the transformations (\ref{newLieonmu}),
(\ref{newLieonB}) are invariances of the action.

Eq.s (\ref{FDAalgebra1})-(\ref{FDAalgebra3}) give the algebra of
diffeomorphisms on the soft FDA manifold.
 \sk
  {\sl Note: } all the variations under diffeomorphisms
  (\ref{diffgroup}),(\ref{diffB2}), (\ref{newLieonmu}),
  (\ref{newLieonB}) can be synthetically written as:
  \eq
  \de \mu^I = (\nabla \epsi)^I + i_{\epsi^J \tbo_J} R^I
  \label{FDAdiff}
  \en
  \noi where $\mu^I = \mu^A, B^i$ etc. If the curvature $R^I$ is
   {\sl horizontal} in some directions {\small {\sl J}} (i.e. if
  $ i_{\tbo_J} R^I =0$), the diffeomorphisms in these directions become
  {\sl gauge transformations}, as evident from (\ref{FDAdiff}).
   In this case a finite gauge transformation can remove the
   dependence on the $y^J$ coordinates, and the fields live on a
   subspace of the original FDA manifold. This generalizes
   horizontality of the curvatures on soft group manifolds: a
   classic example is the Poincar\'e group manifold, where
   horizontality in the Lorentz directions implies Lorentz gauge invariance
    and independence of the fields
   on the Lorentz coordinates.

\subsection{Lie algebra dual of the FDA}

 {} From the algebra of diffeomorphisms (\ref{FDAalgebra1})-(\ref{FDAalgebra3})
 we find the commutators of the Lie derivatives on the rigid
FDA manifold by taking vanishing curvatures and constant $\epsi$
parameters (nonvanishing only for given directions):
 \eqa & &[ \ell_{\tv_A},
\ell_{\tv_B}] = \c{AB}{C} \ell_{\tv_C} + {2 \over p!}
~\cl{ABA_1...A_{p-1}}{i} ~ \ell_{\sigma^{A_1}...\sigma^{A_{p-1}}
\tv_i} \nonumber\\
 & &[\ell_{\tv_A},
\ell_{\si^{B_1}...\si^{B_{p-1}} \tbo_i}] = [\c{Ai}{k}
\de^{B_1...B_{p-1}}_{C_1...C_{p-1}} - (p-1) \c{AC_{1}}{[B_1}
\de^{B_2...B_{p-1}]}_{C_2...C_{p-1}}\de^k_i]
\ell_{\si^{C_1}...\si^{C_{p-1}} \tbo_k} \nonumber\\ &
&[\ell_{\si^{A_1}...\si^{A_{p-1}}
\tbo_i},\ell_{\si^{B_1}...\si^{B_{p-1}}\tbo_j}]=0
 \ena
 \noi This Lie algebra can be considered the dual of the FDA system given in
(\ref{fdaLie}), (\ref{fdaB}), and extends the Lie algebra of
ordinary Lie derivatives (generating usual diffeomorphisms on the
group manifold $G$). Notice the essential presence of the
$(p-1)$-form $\si^A_1...\si^{A_{p-1}}$ in front of the ``tangent
vectors" $\tbo_i$.

\sect{The FDA of D=11 supergravity and its dual}

We recall the FDA of D=11 supergravity \cite{fda1}:
\eqa
 & & d\om^{ab}-\om^{ac} \om^{cb}=0~~~[=R^{ab}]\nonumber\\
 & & dV^a-\om^{ab}V^b-{i\over 2}\psib \Gamma^a \psi =0~~~[=R^a]\nonumber\\
 & & d\psi - {1\over 4} \om^{ab} \Gamma^{ab} \psi = 0 ~~~[=\rho]\nonumber\\
 & & dA - {1\over 2} \psib \Gamma^{ab} \psi V^a V^b
 =0~~~[=R(A)]\label{FDAd11}
 \ena

 \noi The D=11 Fierz identity $\psib \Gamma^{ab}\psi\psib \Gamma^a \psi =0$
 ensures the FDA closure ($d^2=0$). Its Lie algebra part is
 the D =11 superPoincar\'e algebra, whose fundamental fields (corresponding to the Lie algebra generators
 $P_a, J_{ab}, Q$)
 are the vielbein $V^a$, the spin connection $\om^{ab}$ and the gravitino $\psi$. The
 3-form $A$ is in the identity representation of the Lie algebra,
 and thus no $i$-indices are needed.
 The structure constants $\cl{A_1...A_{p+1}}{i}$ of (\ref{fdaB})
 are in the present case given by $\cl{\al\be a b }{}=-12
 (C\Gamma_{ab})_{\al\be}$ (no upper index $i$),
 while the $\c{Aj}{i}$ vanish.

The eq.s of motion on the ``FDA manifold" have the following
solution for the curvatures \cite{fda1}:
 \eqa
 R^{ab}&=& R^{ab}_{~~cd} V^c V^d
 +i~(2\rhob_{c[a}\Gamma_{b]}-\rho_{ab}\Gamma_c)~\psi V^c \nonumber \\
 & & ~~~~~~~~+
 F^{abcd} ~\psib \Gamma^{cd} \psi+ {1 \over 24} F^{c_1c_2c_3c_4}~
 \psib \Gamma^{abc_1c_2c_3c_4} \psi \label{Rab}\\
  R^a &=& 0 \label{Ra}\\
 \rho & =& \rho_{ab} V^a V^b + {i \over 3} (F^{ab_1b_2b_3}
 \Gamma^{b_1b_2b_3}-{1\over 8}F^{b_1b_2b_3b_4}
 \Gamma^{ab_1b_2b_3b_4})~ \psi V^a \label{rho}\\
 R(A)&=& F^{a_1...a_4}V^{a_1}V^{a_2}V^{a_3}V^{a_4}\label{RA}
 \ena

 \noi where the spacetime components $R^{ab}_{~~cd},\rho_{ab},  F^{a_1...a_4}$
 of the curvatures satisfy the well known
  propagation equations (Einstein, gravitino and Maxwell equations):
 \eqa
 & & R^{ac}_{~~bc} - \unmezzo \de^a_b R = 3~ F^{ac_1c_2c_3}
 F^{bc_1c_2c_3}-{3 \over 8}~\de^a_b~F^{c_1...c_4}
 F^{c_1...c_4}\\
 & & \Gamma^{abc} \rho_{bc}=0\\
 & & \Dcal_a F^{ab_1b_2b_3} - {1 \over 2 \cdot 4!\cdot 7!}~
 \epsilon^{b_1b_2b_3a_1...a_8} ~F^{a_1...a_4} F^{a_5...a_8} =0
 \ena

 \subsection{The algebra of diffeomorphisms on the FDA manifold}

Using the structure constants extracted from the FDA
(\ref{FDAd11}) in the general formulas (\ref{FDAalgebra1}),
(\ref{FDAalgebra2}), (\ref{FDAalgebra3}) , one easily finds the
complete diffeomorphism algebra of D=11 supergravity on the FDA
manifold. The superstranslation part reads:
 \eqa & &
 \left[ \ell_{ \epsi^a_1 \tbo_a},\ell_{
\epsi^b_2 \tbo_b} \right]~ =~ \ell_{ \left[ \epsi^a_1 \partial_a
\epsi^c_2 - \epsi^a_2 \partial_a \epsi^c_1 \right] \tbo_c}  - 2~
\ell_{\epsi^a_1\epsi^b_2 R^{cd}_{~~ab} \tbo_{cd}} - 4
~\ell_{\epsi^a_1\epsi^b_2 \psib \Gamma^{ab}\psi ~\tbo}\\ & & \left
[ \ell_{ \epsi^\al_1 \tbo_\al},\ell_{ \epsi^\be_2 \tbo_\be} \right
] = - i \ell_{\epsib_1 \Gamma^c \epsi_2 ~\tbo_c} - 2 ~\ell_{
\epsi^\al_1 \epsi^\be_2 ~R^{cd}_{~~\al\be}\tbo_{cd}}-4~ \ell_{
  \epsib \Gamma_{ab} \epsi  V^aV^b ~ \tbo}\\
   & &  \left[ \ell_{ \epsi^a \tbo_a},\ell_{
\epsi^\be \tbo_\be} \right]= \ell_{(\epsi^a \part_a \epsi^\gamma
 -2 \epsi^a \epsi^\beta \rho^\gamma_{~a\be})~\tbo_\gamma}-
 8~\ell_{\epsi^a \epsib \Gamma_{ab} \psi V^b~\tbo}
 \ena

\noi where $R^{cd}_{~~\al\be}$ and $\rho^\gamma_{~a\be}$ are
respectively the $\psi\psi$ and the  the $V\psi$  components of
the curvatures $R^{cd}$ and $\rho$, as given in eq.s (\ref{Rab})
and (\ref{rho}).

The mixed commutators (between ordinary and extended Lie
derivatives) are computed by adapting the general formula
(\ref{FDAalgebra2}) to the case at hand:
 \eq
 \left[ \ell_{ \epsi^A \tbo_A},\ell_{
\epsi \tbo} \right]= \ell_{(\ell_{\epsi^A \tbo_A \epsi}) \tbo}=
\ell_{ [\epsi^A \part_A \epsi_{BC} + 2 \epsi_{AC} \part_B \epsi^A
+ 4\epsi_{AB}\epsi^D (R^A_{~CD}-\unmezzo C^A_{~CD})]\mu^B\mu^C
~\tbo}
\en
\noi where $\mu^A = V^a, \om^{ab}, \psi^\al$ and the two-form
parameter associated to the three-form $A$ is expanded on the
$\mu^A$ basis: $\epsi = \epsi_{AB} \mu^A \mu^B$. For example:
 \eq
 \left[ \ell_{ \epsi^a \tbo_a},\ell_{
\epsi_{cd}V^c V^d \tbo} \right]= \ell_{(\epsi^a \part_a \epsi_{bc}
+ 2 \epsi_{ac} \part_b \epsi^a)V^b V^c \tbo}
 \en

 Finally, commutators between extended Lie derivatives vanish:
 \eq
\left[ \ell_{ \epsi_1 \tbo_a},\ell_{ \epsi_2 \tbo} \right]=0
 \en
\sk
 \noi {\sl Note: } the action of the extended Lie derivative is nontrivial only on
 $A$, where it amounts to a gauge transformation:
  \eq
   \ell_{ \epsi \tbo} A = d \epsi
   \en
 \noi cf. eq. (\ref{newLieonB}), due to horizontality of $R(A)$ in
 the $A$-direction.

 \subsection{The dual Lie algebra}

   Taking constant parameters ($\epsi^B = \de^B_A$ for a fixed {\small {\sl A}},
   $\epsi_{CD}=\de^{AB}_{CD}$ for fixed {\small {\sl A,B}}) and vanishing
   curvatures, the algebra of Lie derivatives given in the preceding paragraph reduces
   to the following Lie algebra:
   \eqa
  & & \left[ P_a,P_b \right]= -(C\Gamma_{ab})_{\al\be} Z^{\al\be} \nonumber \\
  & & \left[ P_a ,Q_\beta \right] = 2 ~(C \Gamma_{ab})_{\al\be}
  Q^{b\al}  \nonumber\\
   & & \left\{ Q_\al,Q_\be \right \}=i (C\Gamma^a)_{\al\be} P_a +
   (C\Gamma_{ab})_{\al\be}Z^{ab} \nonumber\\
  & & \left[ J_{ab},J_{cd} \right]=\eta_{a[c} J_{d]b}-\eta_{b[c}
  J_{d]a} \nonumber\\
  & & \left[ J_{ab},P_{c} \right]= \eta_{c[a} P_{b]} \nonumber\\
  & &  \left[ J_{ab},Q_{\al} \right]=  -{1\over 4}
  (\Gamma_{ab})_{\al \be} Q_\be  \nonumber\\
  & & \left[ J_{ab},Z^{cd} \right]= 2 \de^{[c}_{[a} Z_{b]}^{~d]} \nonumber\\
  & &  \left[ J_{ab},Q^{c\ga} \right]=  \de^{c}_{[a} Q_{b]}^{~\ga} - {1\over 4}
  (\Gamma_{ab})^{\ga \be} Q^{c\be}\\
  & &  \left[ Q_\al, Z^{ab} \right]=  2i (C\Gamma^{[a})_{\al\be} Q^{b]\be}\label{Liedual}
   \ena
  \noi where only the nonvanishing commutators are given. We have used the familiar symbols for the
  Lie algebra generators $P_a,Q_\al,J_{ab}$ rather than the Lie
  derivative symbols
  $\ell_{\tbo_a},\ell_{\tbo_\al},\ell_{\tbo_{ab}}$.  Moreover we have normalized the
  generators corresponding to the extended Lie derivatives as:
  \eq
   Z^{ab}=4~\ell_{V^a V^b \tbo},~~~Q^{a\al}=4~\ell_{V^a \psi^\al \tbo}
    \en
  To be
  precise, one would also find  $[P_a,P_b]= - 4 (C\Gamma_{ab})_{\al\be} \ell_{\psi^\al \psi^\be \tbo} $.
  However,  when all
  curvatures vanish,  the extended Lie
  derivative $\ell_{\psi^\al \psi^\be \tbo}$ has null action on all the FDA fields, (indeed the only nontrivial action
  $\ell_{\psi^\al \psi^\be \tbo} A$ is proportional to the spin
  connection, which vanishes in flat space). Thus we can set
  $Z^{\al\be} = 4~ \ell_{\psi^\al \psi^\be \tbo} = 0$.

   The third line of (\ref{Liedual}) reproduces the superymmetry
   commutations of M-theory in presence of 2-branes.

  Finally, we give the Cartan-Maurer equations of the Lie algebra (\ref{Liedual}):
   \eqa
 & & d\om^{ab}-\om^{ac} \om^{cb}=0~~~[=R^{ab}]\nonumber\\
 & & dV^a-\om^{ab}V^b-{i\over 2}\psib \Gamma^a \psi =0~~~[=R^a]\nonumber\\
 & & d\psi - {1\over 4} \om^{ab} \Gamma^{ab} \psi = 0 ~~~[=\rho]\nonumber\\
 & & dB^{ab} - \om^{ac} B^{cb} + \om^{bc} B^{ca} - \unmezzo \psib
 \Gamma^{ab} \psi = 0 ~~~[ = T^{ab}] \nonumber\\
 & &  d\eta^{a} - \om^{ac} \eta^{c} - {1\over 4} \om^{cd} \Gamma_{cd} \eta^a
 + 2 C \Gamma^{ab}\psi V^b - 2i~ C \Gamma^{c}\psi B^{ac}= 0 ~~~[ =
 \Sigma^a] \label{CMdual}
  \ena

 \noi where the bosonic one-form $B^{ab}$ and spinor vector one-form $\eta^{a\al}$
 correspond to the generators $Z^{ab}$ and
 $Q^{a\al}$. The closure of this algebra (or equivalently the Jacobi identities for the structure
 constants of the Lie algebra (\ref{Liedual})) can be easily checked
 by use of the D=11 Fierz identity:
 \eq
 \Gamma^{ab} \psi \psib \Ga^b \psi - \Gamma^b \psi \psib
 \Gamma^{ab} \psi =0
  \en
\noi the only nontrivial check concerning the $d\eta^a$ equation
in (\ref{CMdual}).

 \sect{Conclusions}

Generalizing the results of a previous paper \cite{fda3}, we have
further developed an understanding of FDA's in terms of ordinary
Lie algebras. In particular, the symmetries gauged by
antisymmetric tensors are generated by the extended Lie
derivatives introduced in Section 2.

The complete diffeomorphism algebra of FDA's containing a $p$-form
has been obtained, both for the soft and rigid FDA. As in ordinary
group manifolds, the diffeomorphism algebra reduces in the rigid
case to a Lie algebra.

We have applied these results to $D=11$ supergravity, and
recovered the symmetry algebra of the theory, including the
symmetries gauged by the three-form field. Taking its rigid limit
yields the Lie algebra of Section 3, containing the
supertranslation generators $P_a$, $Q_\al$, the Lorentz generators
$J_{ab}$, the familiar pseudo-central charge $Z^{ab}$ and an
additional spinor-vector charge $Q^{a\al}$.

If this algebra can be gauged via the usual procedure of ref.s
\cite{fda1,fda2,gm2} (and there is no a priori reason why it
couldn't) the resulting theory would provide a new formulation of
$D=11$ supergravity in terms of the one-form fields associated to
the Lie algebra generators (i.e. vielbein $V^a$, spin connection
$\om^{ab}$, gravitino $\psi$,  bosonic one-form $B^{ab}$,
spinor-vector one-form $\eta^a$).

We should mention that the D' Auria-Fr\'e algebra of \cite{fda1}
has so far resisted attempts to gauge it: a formulation of D=11
supergravity in terms of the superPoincar\'e fields, bosonic
one-forms $B^{ab}, B^{a_1...a_5}$ and an additional spinor $\eta$
still does not exist. Some recent references on this issue (and on
the use of the D' Auria-Fr\'e algebra in M-theory considerations)
can be found in ref.s \cite{DFinMtheory}

\vfill\eject
\end{document}